\documentclass[noshowpacs, twocolumn,preprintnumbers,secnumarabic,amsmath,amssymb,nofootinbib]{revtex4-1}
\usepackage{dcolumn}      
\usepackage{bm}           
\usepackage{graphicx}
\usepackage{natbib}
\usepackage{caption}
\usepackage{wrapfig}
\usepackage{gensymb}
\usepackage{float}
\usepackage{subcaption}
\usepackage{amsmath,amssymb}  
\usepackage{bm}  
\usepackage[font=small,labelfont=bf]{caption}

\usepackage[pdftex,bookmarks,colorlinks,breaklinks]{hyperref}  
\usepackage{mathrsfs}
\usepackage{amsmath}

\begin{document}
\title{Analogue magnetism revisited}
\author{Bob Osano} 
\email{bob.osano@uct.ac.za} \affiliation{Astrophysics, Cosmology and Gravity Centre, Department of Mathematics and Applied Mathematics, University of Cape Town, Rondebosch 7701, Cape Town, South Africa\\\\}
\affiliation{Academic Development Programme, Science, Centre for Higher Education Development, University of Cape Town, Rondebosch 7701, Cape Town, South Africa}
\author{Patrick W. M. Adams}
\email{pw.adams@uct.ac.za} \affiliation{Astrophysics, Cosmology and Gravity Centre, Department of Mathematics and Applied Mathematics, University of Cape Town, Rondebosch 7701, Cape Town, South Africa\\\\}

\begin{abstract}
In this article we revisit the significance of the often debated structural similarity between the equations of electromagnetism and fluid dynamics. Although the matching of the two sets of equations has successfully been done for non-dissipative forms of the equations, little has been done for cases where the dissipative terms are non-negligible. We consider the consequence of non-negligible viscosity and diffusivity, and how the fine-tuning of these parameters could allow fluid dynamics to be used to indirectly study certain properties of magnetic fields.   
\end{abstract}
\date{\today}
\maketitle
\section{Introduction}
Analogues have made, and continue to make, significant contributions to our understanding of the physics and mathematics of various physical phenomena. These often introduce new ways of looking at the problem at hand, given their ability to allow the cross-fertilisation of ideas from other fields. The task of determining a suitable analogy is a difficult one and and may qualify as a scientific study in its own right. Nevertheless, a carefully chosen analogy can be extremely useful in drawing attention to a specific problem, and the methodologies applied in the process could lead to new and unexpected routes to possible solutions. We are interested in \textbf{the possible analogy between} fluid dynamics and electromagnetism, to which end we ask: to what extent can analysis in fluid dynamics help us understand electromagnetism?

We first need to distinguish two notions that may be the source of confusion in available literature. 1) Matching of variables and equations, which is often on a mathematical level, may be indicative of the possibility of an analogue. The attempts at matching electromagnetism to fluid dynamics goes back to Maxwell \cite{Max,Siegel} who noticed that the magnetic vector potential {\bf A} had characteristics of a moving medium that mimicked the velocity of the flow engulfing a magnetic field line. Today we know that, subject to certain conditions, the following matching holds:
\begin{center}
\begin{tabular}{|l|}
  \hline
{Fluid dynamics$\rightarrow$Electromagnetism} \\
  \hline
Fluid vorticity $\rightarrow$ Magnetic field\\
Fluid velocity $\rightarrow$ Vector potential\\
Fluid acceleration$\rightarrow$Electric field \\
Fluid mass $\rightarrow$Electric charge\\
Fluid density $\rightarrow$ Charge density\\
  \hline
\end{tabular}
\end{center} 
This list is incomplete; for a comprehensive list, the reader is referred to \cite{Martins}.

It is important to keep in mind that the matching of equations and variables is not a sufficient condition to conclude that one has a viable analogue. 2) A successful analogue will, in addition to matching, require similarities in physical behaviour. This points to the need for simulations or experiments. 

 Several authors have gone beyond the matching and have analysed physical behaviours. The authors of \cite{Martins}, compare the inertia forces in fluid dynamics with the inertia forces in electromagnetism. They compare the inertia property of matter in electromagnetism and in the hydrodynamic drag in potential flow, and have found a parallel between the two. The same treatment of the subject is seen in \cite{Arb} where a comparison is made between what the author calls a hydro-electric field and the local acceleration of the fluid, on the one hand, with the Lorentz gauge being compared to the incompressible fluid condition on the other. The author suggests that the Lorentz force in electromagnetism corresponds to the Euler force in fluids. In \cite{Mar} a comparison is made between the inhomogeneous Maxwell equations and two equations encountered in turbulent hydrodynamics, and an analogue to the Poynting vector is proposed. The author of \cite{Wang} makes the comparison between the two fields by introducing Clifford algebras. This is again on a mathematical level and does not delve into the physics. Sound waves within the fluid are compared to the electromagnetic waves in \cite{Tembe} for the case where phase speeds of both waves are independent of wave lengths or non-dispersive. The authors of \cite{cook} point out that the electromagnetic wave equations in a moving medium may be approximated by a form similar to that of the Schr{\"o}dinger equation for a particle in an electromagnetic field, with the velocity and vorticity of the medium playing the roles of the vector potential and magnetic field respectively. An analogue of the Aharonov--Bohm effect is preseneted. Commenting on their simulation, the authors of  \cite{Kulsrud} observe that the root mean square (rms) magnetic field seems to saturate at the same time, and with the same amplitude, as the vorticity does. This is suggestive of a similarity at a physically fundamental level. Other authors (\cite{Kirk}, \cite{Bel},\cite{Leo}, \cite{Rous}, \cite{Siegel},\cite{MarPin}, \cite{Arb}), have also revisited the comparison and have demonstrated, to varying degrees, that the structural similarity goes beyond the vector potential and the velocity.  

These mathematical and physical similarities suggest that fluid dynamics could play the role of an analogue to magnetic fields in the experimental arena. We are interested in two aspects in this study: How fluid dynamics compares to electromagnetism in cases where dissipative terms are present, and where a Biermann battery mechanism operates. These two aspects encode one way in which magnetic field or vorticity are generated, and how they grow. It has been argued \cite{Kulsrud}, that during the pre-galactic era, the magnetic field goes through at least three phases. First, thermo-electric currents generated by shocks increase the magnetic field strength to a value of order $10^{-21}$G, via the Biermann battery mechanism. In second phase, a dynamo action of the turbulence operates faster than the battery action; a phase which sees the turbulence formation of a Kolmogorov spectrum down to the viscous scale.  It is thought that the smallest eddy  drives amplification of the magnetic field and continues to amplify it until saturation is encountered. The third phase sets in when saturation is approached. In this phase, the magnetic field approaches equipartition with the hydrodynamic turbulence and becomes coherent on very large scales. 

Our analysis considers the first and the second phases for a possible shock-generated magnetic field, and compares to analogous behaviour in fluid dynamics. The analysis of turbulence requires an understanding of the Reynolds number, and hence, the kinetic viscosity. It will be appreciated that despite the enomous work that has been done, a complete theory of MHD turbulence remains elusive. Such important questions as: 1) What role does turbulence play in the amplification, the preservation and the shape magnetic fields, and, 2) What is the structure and spectrum of such a field at different scales, are still unclear. Turbulence \cite{sche} is therefore not only important in astrophysics, but also in the problem of magneto-genesis. The study of analogues may just provide clues on how to deal with the challenges encountered this far.

\section{Magnetic induction equation and the Vorticity equation for viscous fluids}
We will compare two types of fluids: a magnetic fluid $\mathcal {MF}$ (approximated as a fluid or simply a non ideal MHD) and a viscous fluid $\mathcal{VF} $. The two are examined separately and compared. The effects of entropy are neglected. Let us start with the magnetic field.

\subsection{Magnetic fluid}
In the most general case, magnetohydrodynamics is the idealisation of a compressible fluid that is made up of point charges that are moving in the presence of a magnetic field \cite{Walker}, thus making the fluid to be a conducting medium. The fluid in this case is composed of a comparable number of negatively- and positively-charged particles. The behaviour of such a fluid can be approximated using the laws of thermodynamics and kinetic theory. In particular, the properties of these moving particles are averaged over small enough volumes in comparison to the macroscopic volumes, but at the same time sufficiently large in comparison to the distances between particles. 

Plasma that is fully ionised may be viewed as a system of point charges in motion. In this case, the charge density, {\it q}, and the current density, {\bf J}, which are functions of position and time, determine both the positions and velocities of these charges. Assuming that the particles are not bound, and that the magnetic properties arising from the orbital and spin angular momentum can be neglected, then the electromagnetic field in such a medium may be described by the two field variables that are appropriately named; {\bf B} (magnetic) and {\bf E} (electric). On the scales of individual particles, the charge and current densities fluctuate wildly exhibiting $\delta$-function behaviour; they appear as zero everywhere except at the location of the particles. Considering a single particle located at position {\bf r}, having a charge value {\it q} and velocity {\bf v}. The charge density $\eta$ is given by:

\begin{eqnarray}
\eta({\bf r})&=&q\delta^3({\bf r})
\end{eqnarray} and when the charge is located in a volume {\it V}, 
\begin{eqnarray}
q &=& \int_{V} \eta({\bf r}) dV.
\end{eqnarray} The average charge in a volume {\it V} is then given by
\begin{eqnarray}
\langle \eta \rangle &=& \frac{1}{V}\int_{V} \eta({\bf r}) dV.
\end{eqnarray} The angle brackets imply average over volume in this context. In theory, the fluid approximation is considered as the limit approached when {\it V} is made considerably small, but large compared to inter-particle distances. Similarly, one can define the average current density $\langle {\bf J}\rangle$. Given these length scales, the actual particle distribution may be approximated by the average particle distribution. It is important to note that if the volume is further reduced to a size comparable to the inter-particle distances, then fluid approximation breaks down. The determination of such limits is no trivial matter and is something worth investigating. 

\subsection{Comments on coarse-graining and fluid approximation}
Let us consider what we mean by conservation of particles given an arbitrary volume, {\it V}, having a surface area, {\it S}. We require that the rate of increase in the number of particles in this volume, must of necessity be minus the rate at which the particles flow out through the surface. Hence,
\begin{eqnarray}
\label{eq:7}\frac{d}{d t} \int_{V} NdV = - \oint_{S}N\langle{\bf v}\rangle\bold{.}d\bold{S},
\end{eqnarray}
where $N$(= $n/V$, if n is the actual number of particles) is the number density and $\langle \bold{v}\rangle$ is the average velocity. $d/dt$ operates on the particles which moves with the average fluid velocity. At a point in space,
\begin{eqnarray}
\label{eq:8}\frac{d}{d t}  =  \frac{\partial }{\partial t}+ \langle{\bf v}\rangle\bold{.}\nabla.
\end{eqnarray}
Using divergence theorem, Eq. (\ref{eq:7}) can be written in the form
\begin{eqnarray}
\label{eq:9}\int_{V}\left\{ \frac{\partial N}{\partial t} + \nabla\bold{.}(N\langle{\bf v}) dV\right\}=0, 
\end{eqnarray} which is the equivalent of Euler's equation for the conservation of mass.
Eq. (\ref{eq:8}) is clearly about the number of discrete particles in $V$, while Eq. (\ref{eq:9}) is the fluid continuity equation. Hidden in this mathematical formulation and physical assumptions is the underlying transition from discrete to continuum behaviour, as is well known in continuum mechanics. What is not so well known is the averaging problem, in which the time derivative and averaging procedures may not commute. This shows up when we consider the momentum equation for example. Consider the case where particles move with the average velocity such that there are no thermal motions. The bulk behaviour is given by the Lorentz force multiplied by the number density:
\begin{eqnarray}
\label{eq:10} mN\frac{d \langle {\bf v}\rangle}{d t}  = qN\{{\bf E}+ {\langle {\bf v}\rangle}\times{\bf B}\},
\end{eqnarray} where $m$ is the mass. It should be borne in mind that although $\frac{d{\langle{\bf v}\rangle}}{d t}$ is equal to $\langle \frac{d{\bf v}}{d t}\rangle$ for incompressible fluids, the equality does not hold in general. This issue will be addressed elsewhere \cite{bob2016}\\.

Two issues standout in this narrative. On one hand, we have the issue of the extent to which our fluid approximation holds ($\mathcal{MF}$) and on the other hand we have the issue of determining suitable fluid ($\mathcal{VF}$) for comparative purposes. These two aspects are important in determining suitable analogies. Before we attempt to do this though, it is important to lay down the foundation for the rest of the study. On the scales for which fluid approximations are valid, one can derive the induction equation:
\begin{eqnarray}
\label{eq:1}\frac{\partial{\bold{B}}}{\partial{t}}-\nabla\times(\bold{v}\times\bold{B})=0,
\end{eqnarray}
where $\bold{B}$ is the magnetic flux and $\bold{v}$ is the velocity of conducting fluid. Starting with Maxwell's equations and Amp{\`e}re's law, one can show that as long as the rate at which the electric field varies exceeds Faraday time, then the displacement current can be eliminated from the system of equations\cite{Axel}, leading to equation Eq.(\ref{eq:1}) gaining a diffusive term;
\begin{eqnarray}
\label{eq:2}\frac{\partial{\bold{B}}}{\partial{t}}-\nabla\times(\bold{v}\times\bold{B}-\eta\nabla\times\bold{B})&=&0,
\end{eqnarray}
where $\eta$ is the coefficient of diffusion (diffusivity). In effect, the presence of the diffusive term signifies a {\it non-ideal MHD} consideration. It is important to note the implication of equations Eq.(\ref{eq:2}) and Eq.(\ref{eq:1}): namely that ${\bf B}=0 $ is a solution of the two. Physically this means that if the magnetic flux is initially zero, it will remain zero.

The difference between equations Eq.(\ref{eq:1}) and Eq.(\ref{eq:2}) is the diffusivity term in the latter. This term is often set to zero \cite{Kulsrud} when the induction equation is compared to the viscosity equation for non-viscous fluids. We would like to examine the similarity between the two fluids at a deeper level and will therefore seek to maintain the diffusivity term. If we consider plasma fluid, as previously discussed, electrons are generally accelerated relative to ions in the presence of an electric field. The motion of these electrons will be hampered by frictional forces that arise from electron-ion collisions. This leads to the curtailed velocity of electrons, relative to the ions, and which results from the act of balancing the Lorentz force with the friction force. This accounts for the diffusion. The measure of the velocity of electrons relative to ions can be found using the common assumption \cite{Axel} that electrons move freely for about an electron-ion collision time, after which their velocity again becomes randomised. We now look at the viscous fluid ($\mathcal{VF}$).

\subsection{Viscous Fluid}
Helmholtz laws, which are conservation principles that allow for a consistent analysis of the evolution of vorticity, require among other things that the viscosity be negligible. Fluids governed by these laws are idealised. In order to study a more realistic fluid, we require a non-negligible viscosity -- in effect, a violation of Helmholtz's restriction. The vorticity equation that we consider in this section is that of a non-charged viscous fluid.

 The motion of an incompressible viscous fluid, with no external body forces other than viscous forces, is modelled by the Navier-Stokes equation:
\begin{eqnarray}
\label{eq:5}\frac{\partial{\bold{u}}}{\partial{t}}+\bold{u}\bold{.}\nabla{\bold{u}}+\frac{\nabla p}{\rho}-\nu\nabla^2\bold{u}=0.
\end{eqnarray} Note that one easily recovers the Euler conservation of momentum equation when the viscosity, $\nu$, is set to zero. As is the case for a high Reynolds number (Re$\gg$1).
The Reynolds number characterises the magnitude of inertial effects compared to the magnitude of viscous effects. A low Reynolds number (Re$\ll$1) shows that viscous forces in the fluid are very strong compared to inertial forces.

We take the curl of Eq.(\ref{eq:5}), employing the identity
\[\bold{u}\bold{.}\nabla\bold{u}=\nabla(\frac{1}{2}\bold{u}^2)-\bold{u}\times(\nabla\times\bold{u}),\]
and the fact that curl commutes with the Laplacian operator i.e. $curl\nabla^2=\nabla^2 curl$. This gives the vorticity equation:
\begin{eqnarray}
\label{eq:6}\frac{\partial{\bold{\omega}}}{\partial{t}}=\nabla\times(\bold{u}\times\bold{\omega}-\nu\nabla\times\bold{\omega})
\end{eqnarray}
where $\omega= \nabla\times\bold{u}$ is the vorticity. Eq.(\ref{eq:6}) is similar to Eq. (\ref{eq:2}), save for the the diffusive and the viscous term. In fact if these two terms are neglected, one can write a single equation:
\begin{eqnarray}
\frac{\partial{\bold{\mathcal{D}}}}{\partial{t}}-\nabla\times(\bold{\mathcal{U}}\times\bold{\mathcal{D}})&=&0,
\end{eqnarray} where ${\bf \mathcal{D}}$ is either ${\bf B}$ or ${\bf -\omega}$ and $\mathcal{U}$ is either ${\bf u}$ or ${\bf v}$. The reason for the negative vorticity will become apparent when consider the presence of the Biermann battery term ( which arises as an extra pressure gradient term in Ohm's law \cite{Bier}): The similarity between the two, with each bearing a battery term, was pointed out  in \cite{Kulsrud}. \begin{eqnarray}
\label{eq:4a}\frac{\partial{\bold{B}}}{\partial{t}}-\nabla\times(\bold{v}\times\bold{B}-\eta\nabla\times\bold{B} )=\frac{c\nabla p_{e}\times\nabla n_{e}}{n_{e}^2 e(1+\chi)},\\
\label{eq:4b}\frac{\partial{\bold{\omega}}}{\partial{t}}-\nabla\times(\bold{u}\times\bold{\omega}-\nu\nabla\times\bold{\omega} )= - \frac{c\nabla p \times\nabla\rho}{\rho^2 }.
\end{eqnarray} where $e$ is the charge, $p_{e}$ the electron fluid pressure, $\rho_{e}$ the electron number density and $\chi$ is the ionisation fraction. The speed of light, ${\it c}$, could be consistently set to unity. As pointed out by \cite{Axel}, if one assumes that the charged fluid is made up of free electrons, protons ( ionised hydrogen) and hydrogen atoms, with ionisation fraction $\chi$ (constant in space), and the same temperature for the all species present, then it is possible to recast the equation for $p_{e}=\chi p/(1+\chi)$  and $n_{e}=\chi\rho/m_{p}$ is electron density with $m_{p}$ the proton mass. This makes the RHS of Eq. (\ref{eq:4a}) to equal the positive of the RHS of Eq. (\ref{eq:4b}). We will take up the case with the battery term elsewhere \cite{bob2016}. For now we extend the comparison to the potentials for both the magnetic flux and the fluid vorticity.

\begin{figure}[htp]
\tiny{ 
\caption{\it The parameters $\eta$ and $\nu$ are set to $0$. The {\it rms} magnetic flux {\bf B} is then plotted against the vorticity $\omega$. We have used Eqs. (\ref{eq:4a}) and (\ref{eq:4b}) without the battery terms.  \label{fig:01}}}
  \centering
  \includegraphics[width=0.35\textwidth]{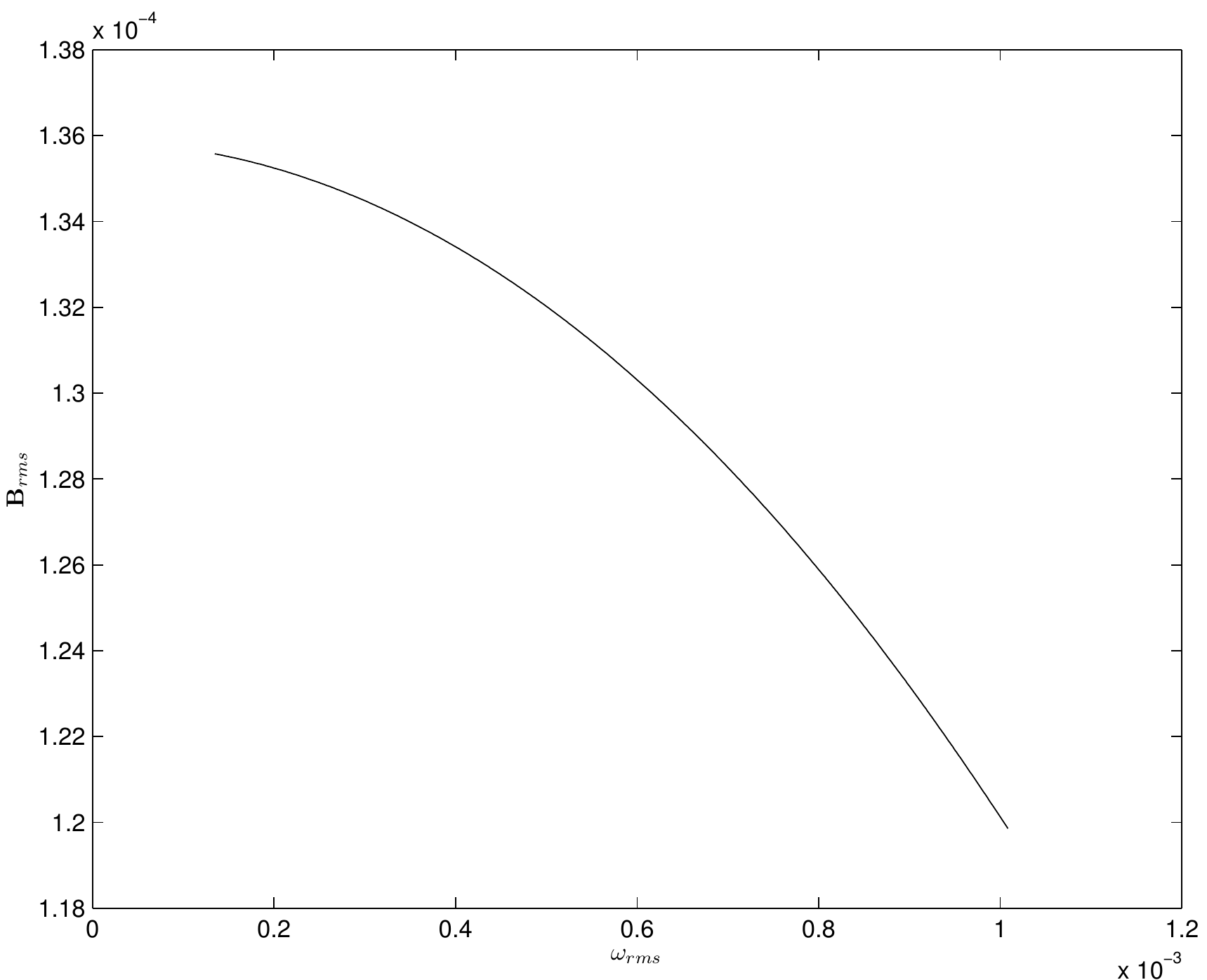}

\end{figure}

We would now like to present the simulation of magnetic field and fluid dynamics. In this regard we use Eqns (\ref{eq:4a}) and (\ref{eq:4b}). The case of a homogeneous flow, with no dissipation is given same initial conditions is given in Figure (\ref{fig:01}). But, as stated before, we are interested in dissipative flows.

Could we compare different kinds of fluids ($\mathcal{VF}$) to the ($\mathcal{MF}$)?  In order to attempt this, we first need to examine kinematics viscosities of some common fluids. These have been obtained from \cite{Landau}, and are quoted at $20\degree$C. 

Simulations were done using the \textsc{Pencil Code}\footnote{\url{http://pencil-code.nordita.org/}}, in which a $32^3$ periodic box of dimensions $2\pi\times2\pi\times2\pi$ was considered. All initial conditions were set to Gaussian noise of small amplitude, and the temporal growth of the rms strengths of the relevant quantities were observed and are presented here.

\begin{table}[h!]
\caption{Kinematic viscosities of some fluids}\label{table1} 
\begin{tabular}{l*{3}{l}c}
Fluid            &~~~~~~~~$\nu(cm^{2}/sec)$\\
\hline
Water &~~~~~~~~$1.0 \times10^{-2} $\\
Air &~~~~~~~~$1.5\times 10^{-1}$\\
Alcohol &~~~~~~~~$2.2 \times 10^{-2}$\\
Mercury &~~~~~~~~$1.2 \times10^{-3}$\\
\hline
\end{tabular}
\end{table}

\begin{figure}[h!]
\tiny{\caption{\it The parameters $\eta=\nu=10^{-2}$; numerically. The magnetic field ${\bf B} _{rms}$ plotted against the vorticity ${\bf \omega}_{rms}$. We have again used Eqs. (\ref{eq:4a}) and (\ref{eq:4b}) without the battery terms.\label{fig:021}}} 
  \centering
  \includegraphics[width=0.35\textwidth]{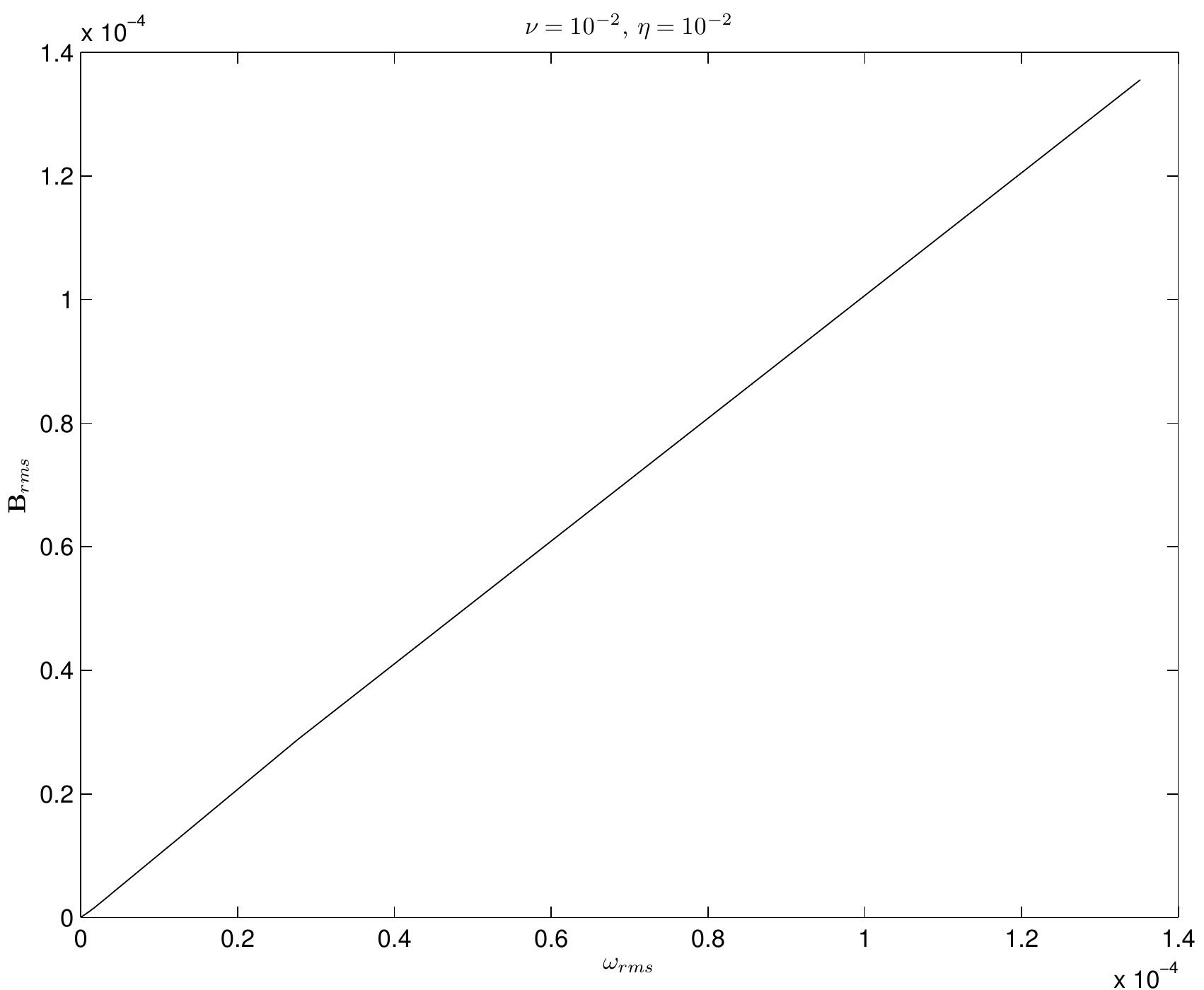}
\end{figure}

The case of comparison to hydrodynamics is given in Figure (\ref{fig:021}).

\section{Evolution equations for magnetic potential and fluid velocity}
Much of the discussion in this section follows that in the previous section. Let us begin with Maxwell's equations, which in SI units take the form:
\begin{eqnarray}
\label{eq:5a}\frac{\partial{\bf B}}{\partial t}&=& - \nabla\times{\bf E},\\
\label{eq:5b}\frac{\partial{\bf E}}{c^{2} \partial t}&=&  \nabla\times{\bf B} -\mu_{0}{\bf J},\\
\nabla{\bf .}{\bf B}&=&0,\\
\nabla{\bf .}{\bf E}&=&\frac{\rho_{e}}{\epsilon_{0}},
\end{eqnarray}

where the variables have the same meaning as previously given, and $\epsilon_{0}=1/(\mu_{0}c^2)$ is the permittivity of free space.
In order to keep $\nabla {\bf {.}~B} = 0$ at all times, it is convenient to write ${\bf B}=\nabla\times{\bf A}$, where ${\bf A}$ is the vector potential for magnetic field. We note that one can give the uncurled form of Eq (\ref{eq:5a})
\begin{eqnarray}
\label{eq:6a}\frac{\partial{\bf A}}{\partial t}&=&-{\bf E}-\nabla\phi,
\end{eqnarray} where $\phi$ is a scalar potential.  Taking the curl of Eq. (\ref{eq:6a}), applying the generalised ohm's law to eliminate {\bf E}, but at the same time keeping the term $\nabla\times\nabla\phi$, although we know it vanishes, yields
\begin{eqnarray}
\label{eq:7a}\nabla\times\frac{\partial{\bold{A}}}{\partial{t}}=\nabla\times(\bold{u}\times\nabla\times\bold{A}-\eta\nabla^2\bold{A}-\nabla\phi),
\end{eqnarray}
where we have used the relation:
\begin{eqnarray}
\nabla\times\nabla\times{\bf A}=\nabla(\nabla{\bf .}{\bf A})-\nabla^2 {\bf A},
\end{eqnarray} and the condition $\nabla{\bf . }{\bf A}=0.$ Note that if the curl is dropped from both sides of Eq.(\ref{eq:7a}), one gets
\begin{eqnarray}
\label{eq:7b}\frac{\partial{\bold{A}}}{\partial{t}}=(\bold{u}\times\nabla\times\bold{A}-\eta\nabla^2\bold{A}-\nabla\phi),
\end{eqnarray} which is similar to Eq.(\ref{eq:5a})
Comparing the evolution equation of the magnetic potential $\bold{A}$ to the Naiver-Stokes equation.
\begin{eqnarray}
\label{eq:8a}\frac{\partial{\bf A}}{\partial t}&=&{\bf u}\times(\nabla\times{\bf A})+\eta\nabla^2 {\bf A}-\frac{\nabla p_{e}}{n_{e}}-\nabla(\phi)\\
\label{eq:8b}\frac{\partial{\bold{u}}}{\partial{t}}&=&{\bf u}\times(\nabla\times{\bf u})+\nu\nabla^2\bold{u}-\frac{\nabla p}{\rho}-\nabla(\frac{1}{2}{\bf u}^2)
\end{eqnarray}
where $\phi$ is a scalar potential. We note that the two equations look identical and one can compare term by term. Note that we have used Amp{\`e}re's law in place of generalised ohm's law. The two last terms in the above equations can be brought together to read: $ - \nabla (p/\rho+\phi)$ and $ - \nabla (p/\rho+{\bf u}^{2}/2)$, where the latter is the negative of the Bernoulli energy function. An example in hydrodynamics where this is possible is the homogeneous flows ( $\nabla \rho=0).$ In general $\nabla\rho\neq 0$, and so we leave the two as separate terms. Also to be noted is that the similarity holds even when one invokes potential flows \cite{Landau} in hydrodynamics ($\nabla\times {\bf u} = \omega = 0$; {\it Irrotational} flows). This case corresponds to the case {\bf B} =$\nabla\times {\bf A}=0,$ which is of little interest in our context.

\begin{figure}[h!]
\tiny{\label{fig: AandU} 
\caption{\it Shows the plot of {\it rms} magnetic potential ${\bf A}$ against {\it rms} velocity ${\bf u}$, where Eqs. (\ref{eq:8a}) and (\ref{eq:8b}) have been used with the conditions: $\eta=\nu=10^{-2}$. The reason for choosing this condition is given in the appendix section }}
  \centering
  \includegraphics[width=0.35\textwidth]{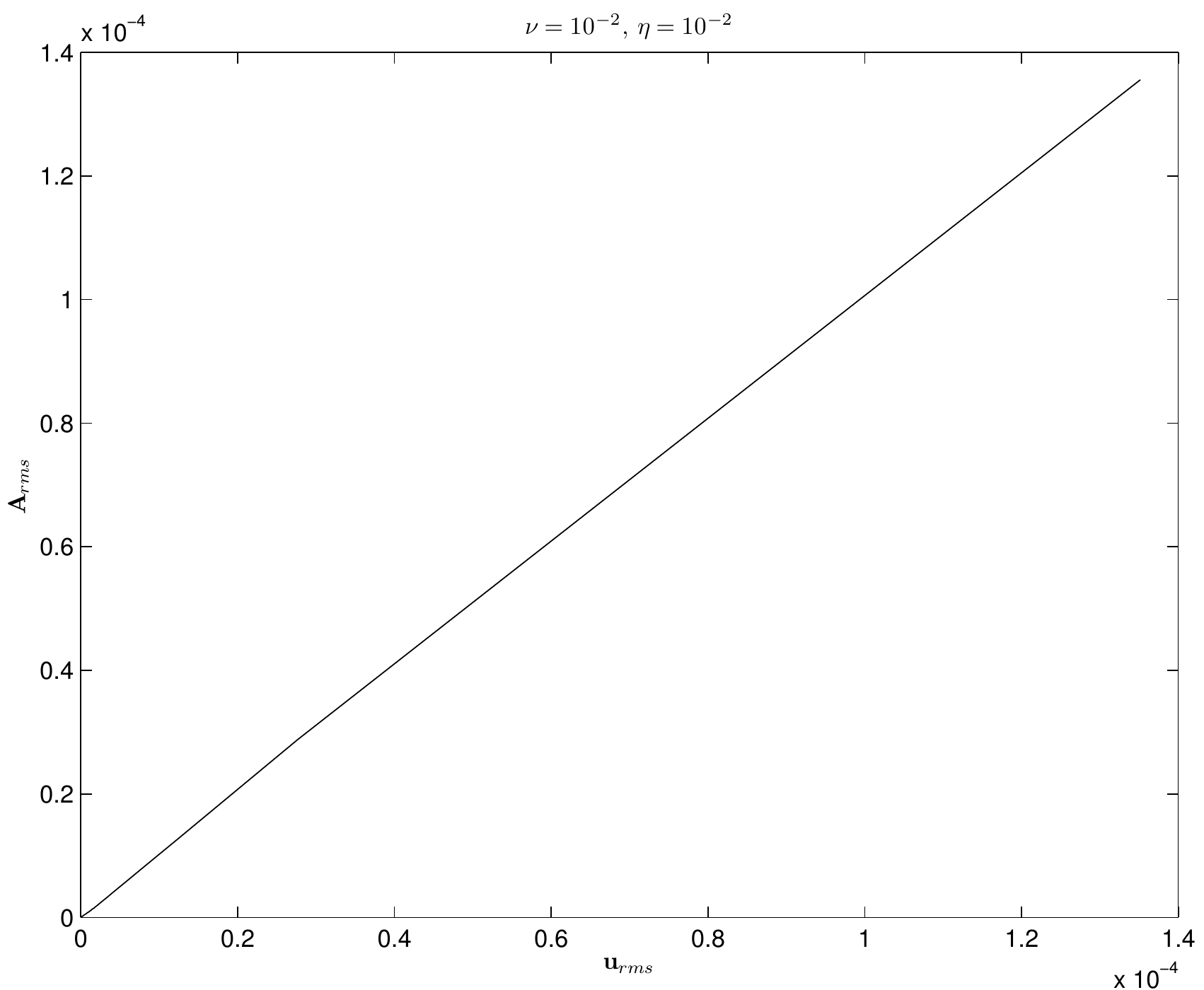}
\end{figure}

\section{Discussions and Conclusion} 
The origin of cosmic magnetic field on large scales is not well understood, despite decades of work that has gone into it. We argue that analogues could play a significant role in making the required advancements. Indeed, analogues are known to have made contributions to deeper understanding of various physical phenomena, via the introduction new ways of looking at the problem at hand and borrowing methodologies from other fields. Magneto-genesis and magnetic field amplification should not be an exception.

It will be appreciated that despite the enomous work that has been done, a complete theory of MHD turbulence remains elusive. Such important questions as: 1) What role does turbulence play in the amplification, the preservation and the shape of magnetic fields? 2) What is the structure and spectrum of such a field at different scales? These are still unclear. Turbulence \cite{sche} is therefore not only important in astrophysics, but also in the problem of magneto-genesis. The study of magnetic analogue models may just provide clues of how to deal with the challenges encountered thus far. The development of magnetic analogue model begins with looking at structural similarities between magnetic field and a suitable analogue; in this case fluid dynamics.

The comparison between equations of fluid dynamics and magnetic field has been made by various authors beginning with Maxwell himself. We have argued in this article that much of the comparison has been at the level of the mathematical structure. We think that the future prospects for magnetic field analogue models look bright. The development of such models will require the comparisons to go beyond the mathematical structures. As a first step, we have looked at viscosity, {\it vis-{\`a}-vis}, diffusivity because of the role these play in the definition of their corresponding Reynolds numbers and hence on the understanding of the onset of turbulence in their respective fluids. We have demonstrated that by fine-tuning both kinematic viscosity (in this case for hydrodynamics) and diffusivity, one obtains similarity in the behaviours of ${\bf B}_{rms}$ and ${\bf \omega}_{rms}$, on the one hand, and ${\bf A}_{rms}$ and ${\bf u}_{rms}$ on the other. There is no a priori reason why this cannot be for done for other fluids, and will be demonstrated in \cite{Patrick}. It has been pointed out elsewhere \cite{sche} that there is a puzzling refusal of the numerical MHD turbulence to agree with with the solar-wind observations, and that this highlights the shaky quality of the existing physical understanding of what really happens in a turbulent magnetic fluid on the dynamical level. This may just benefit from experiments with analogues. 

\section{Acknowledgement}
Patrick W. M. Adams acknowledges funding support from the National Research Foundation (NRF) of South Africa, as well as funding from the University of Cape Town, both administered by the Postgraduate Funding Office (PGFO) of the University of Cape Town. Bob Osano acknowledges URC funding support administered by the University of Cape Town.
 

\section{Appendix}
\appendix

\section{Constraining viscosity and diffusivity}\label{ConstDV}

\begin{widetext}
\subsubsection{Viscosity} 

We are interested in studying the effect of viscosity, {\it vis-{\`a}-vis}, the effect of diffusion. In this section we examine how the velocity decays given different viscosities. As expected the greater the viscosity, the faster the reduction in velocity. Although $\nu=0.00001$ gives the lowest decay rate of the the three non-zero viscosities. Slower reduction in velocity means longer time for visualisation. But $\nu=0.001$ corresponds to hydrodynamics at $20\degree$C and will therefore be of interest. For this case we note that the $32^3$ case shows the slowest decay and is therefore best for the comparative analysis, not to mention that it is cheaper in terms of computing time. 
 \begin{figure}[H]
 \begin{subfigure}{0.5\textwidth}
 \centering
 \includegraphics[scale=0.4]{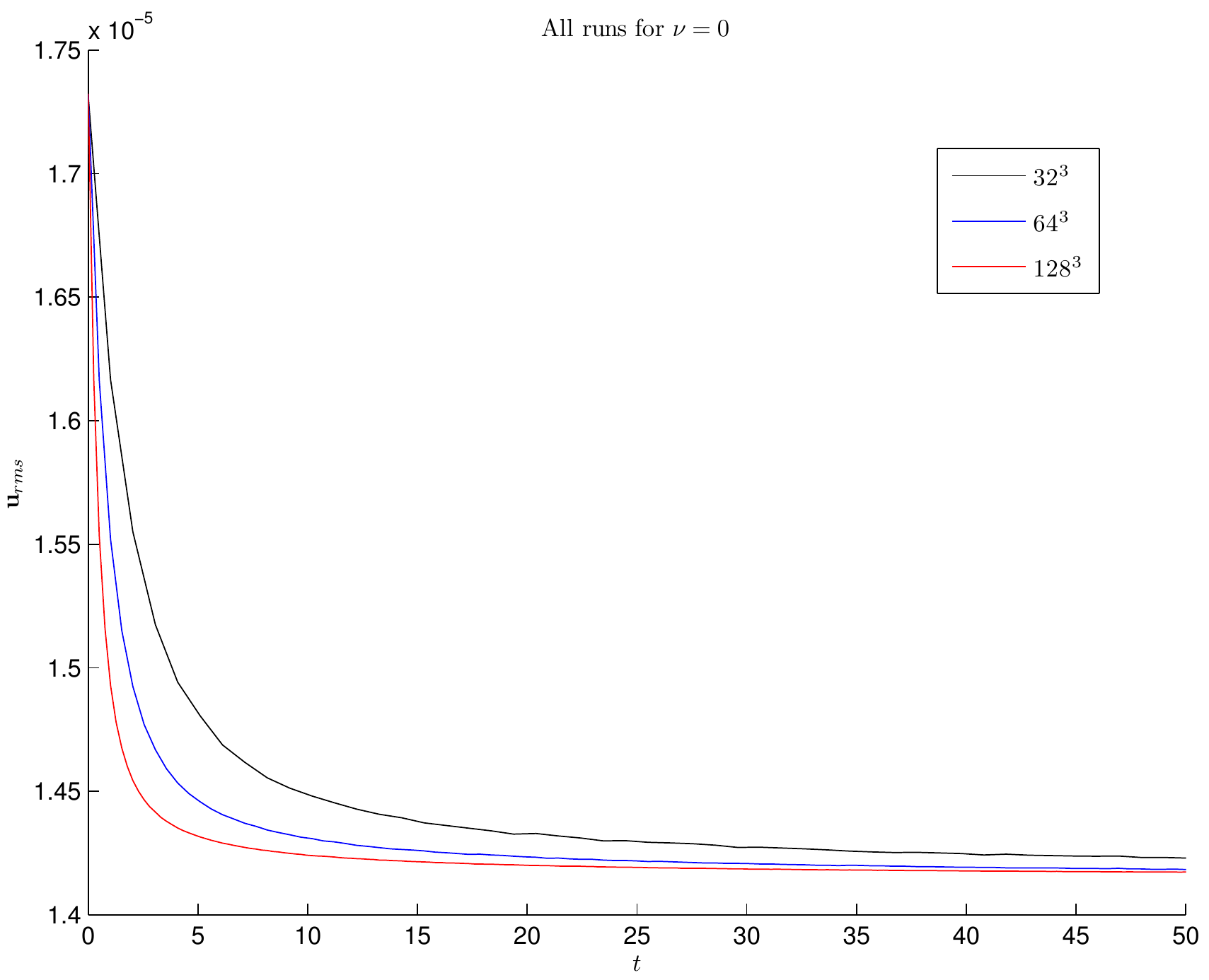}
 \caption{$\nu$=0}
 \label{fig:sub5}
 \end{subfigure}
 \begin{subfigure}{.5\textwidth}
  \centering
  \includegraphics[scale=0.4]{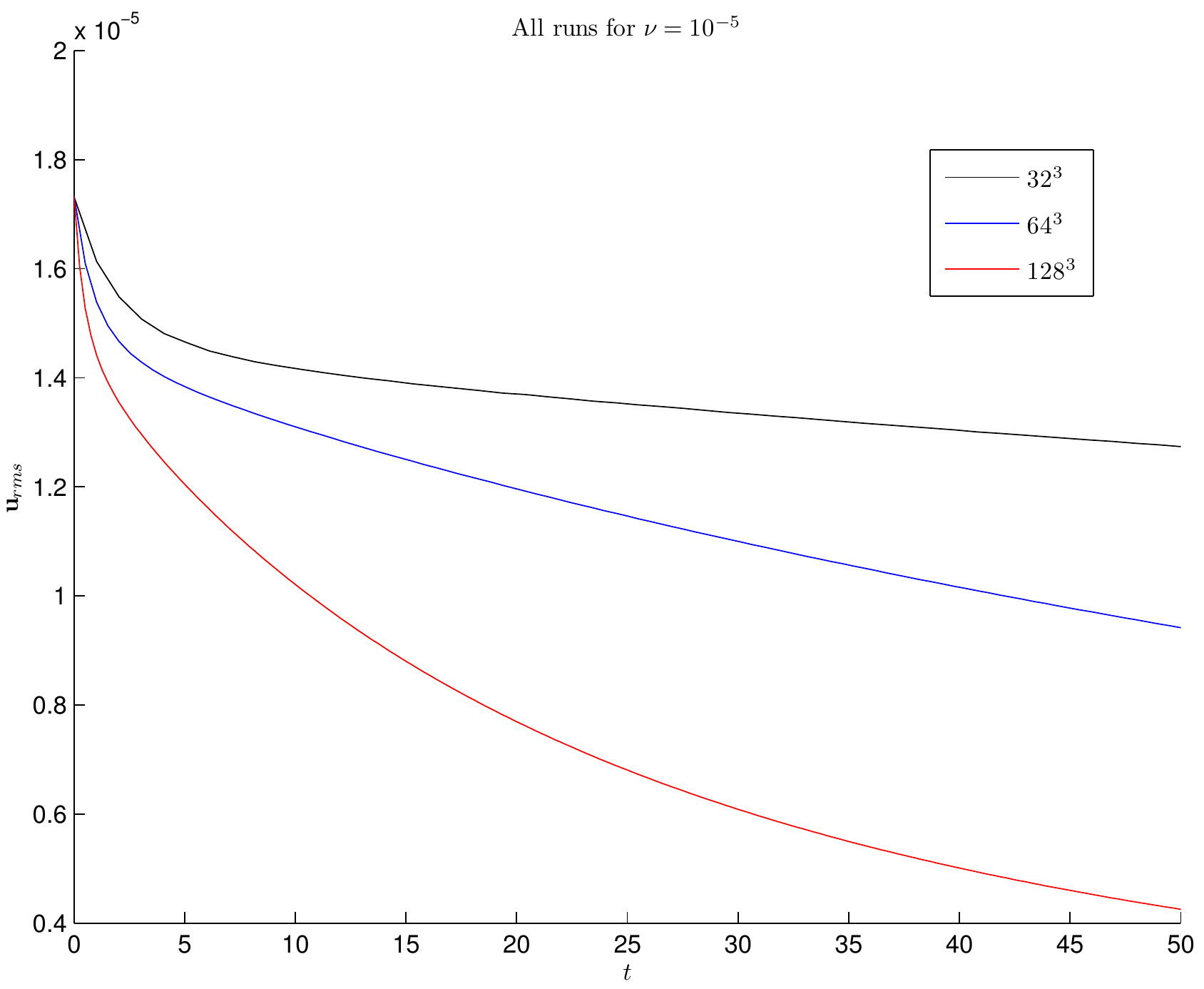}  
 \caption{$\nu$=0.00001}
  \label{fig:sub6}
\end{subfigure}%
\\
\begin{subfigure}{.5\textwidth}
  \centering
  \includegraphics[scale=0.4]{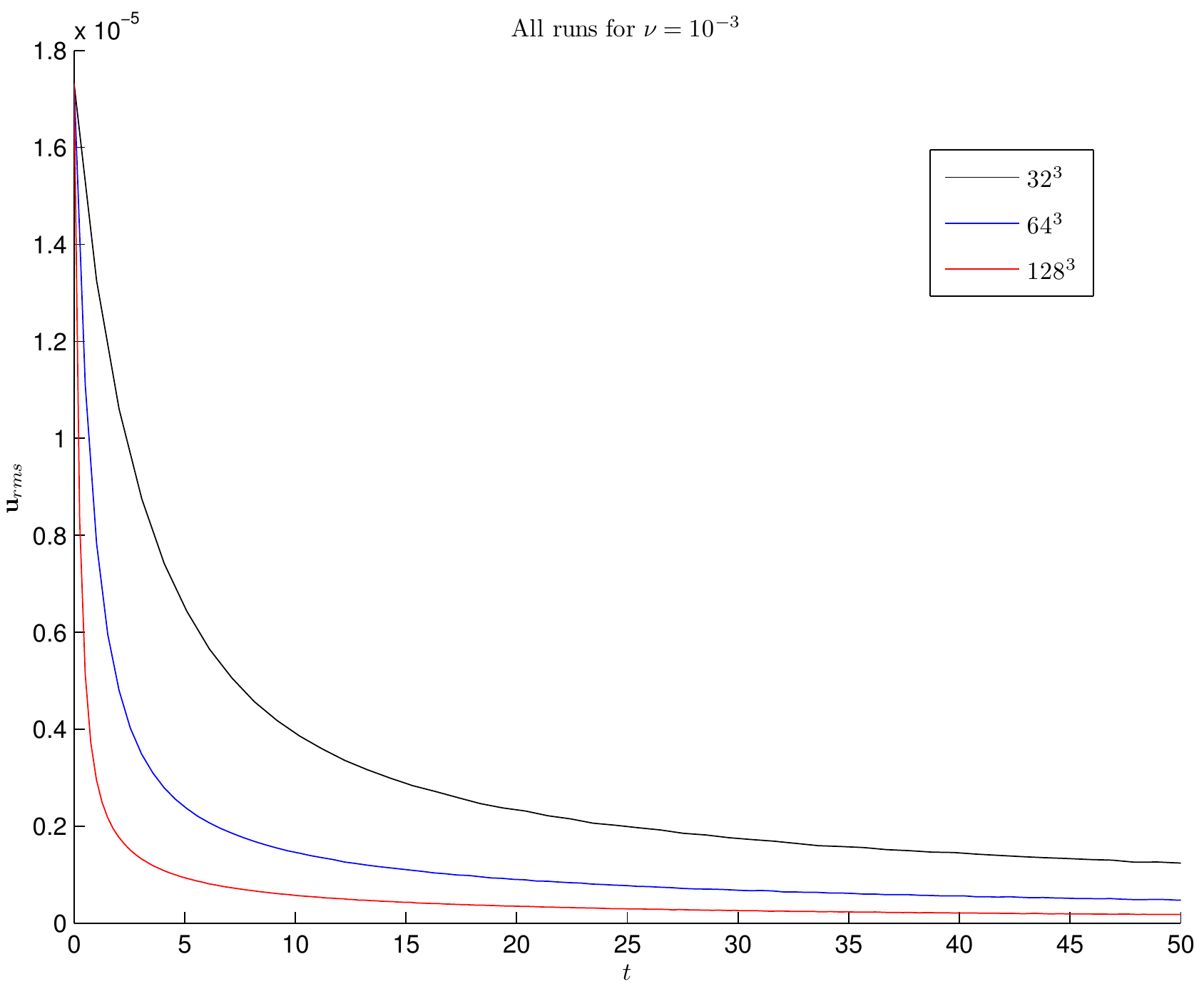}
  \caption{$\nu$=0.001}
  \label{fig:sub7}
\end{subfigure}%
\begin{subfigure}{.5\textwidth}
  \centering
  \includegraphics[scale=0.4]{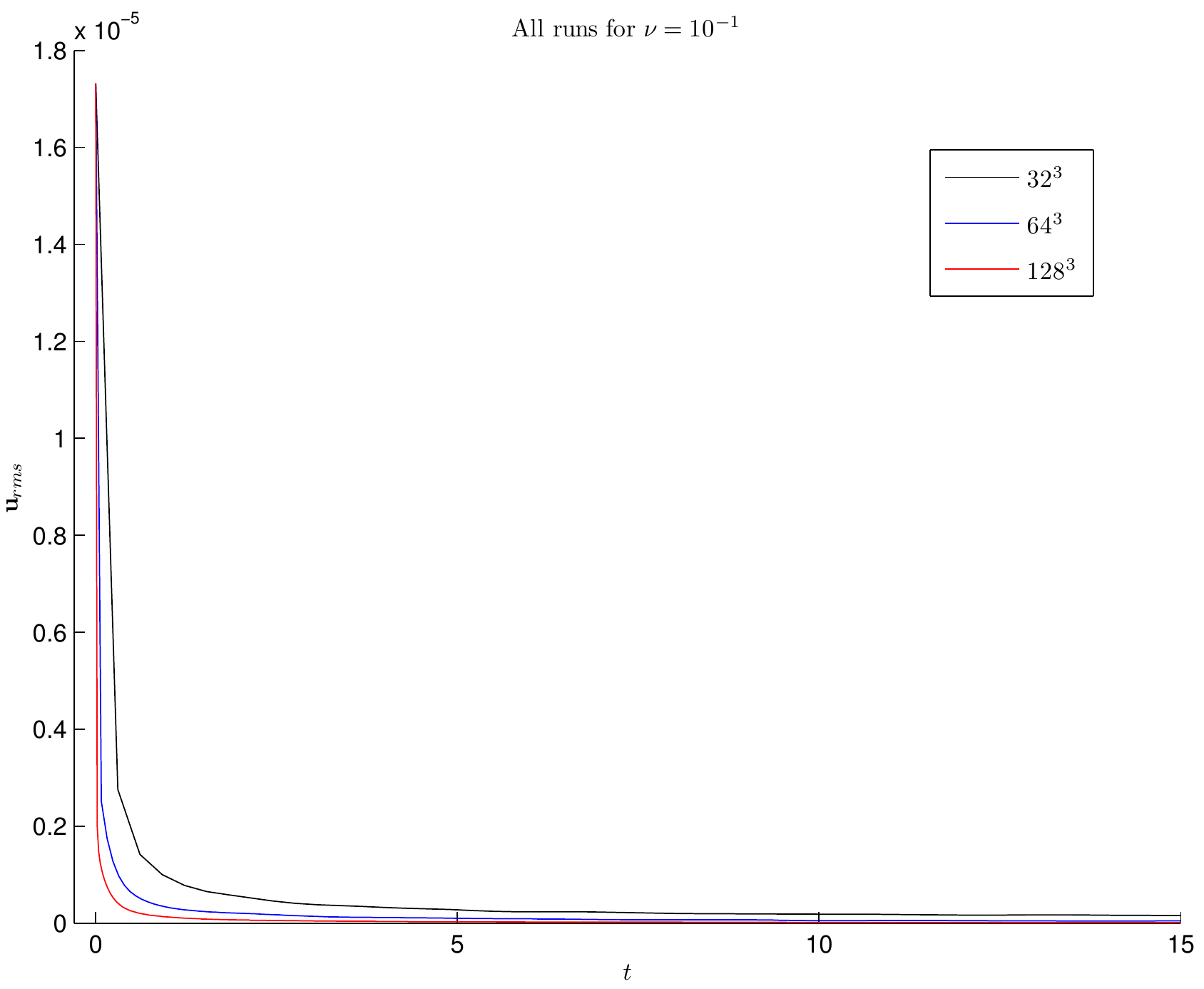}
  \caption{$\nu$=0.1}
  \label{fig:sub8}
\end{subfigure}%
\end{figure}

\subsubsection{Diffusivity}
Similarly, we examine how the magnetic potential decays, given different values of diffusivities. A behaviour similar to velocity is seen here, in particular, the higher the diffusion, the faster the reduction in the magnitude of the magnetic potential. As in the viscosity case, the $32^3$ mesh gives the slowest decay for the potential. The unexpected behaviour in \ref{fig:sub1} for $128^{3}$ is attributed to numerical effects.
\begin{figure}[H]
\label{figD}
\begin{subfigure}{.5\textwidth}
  \centering
  \includegraphics[scale=0.4]{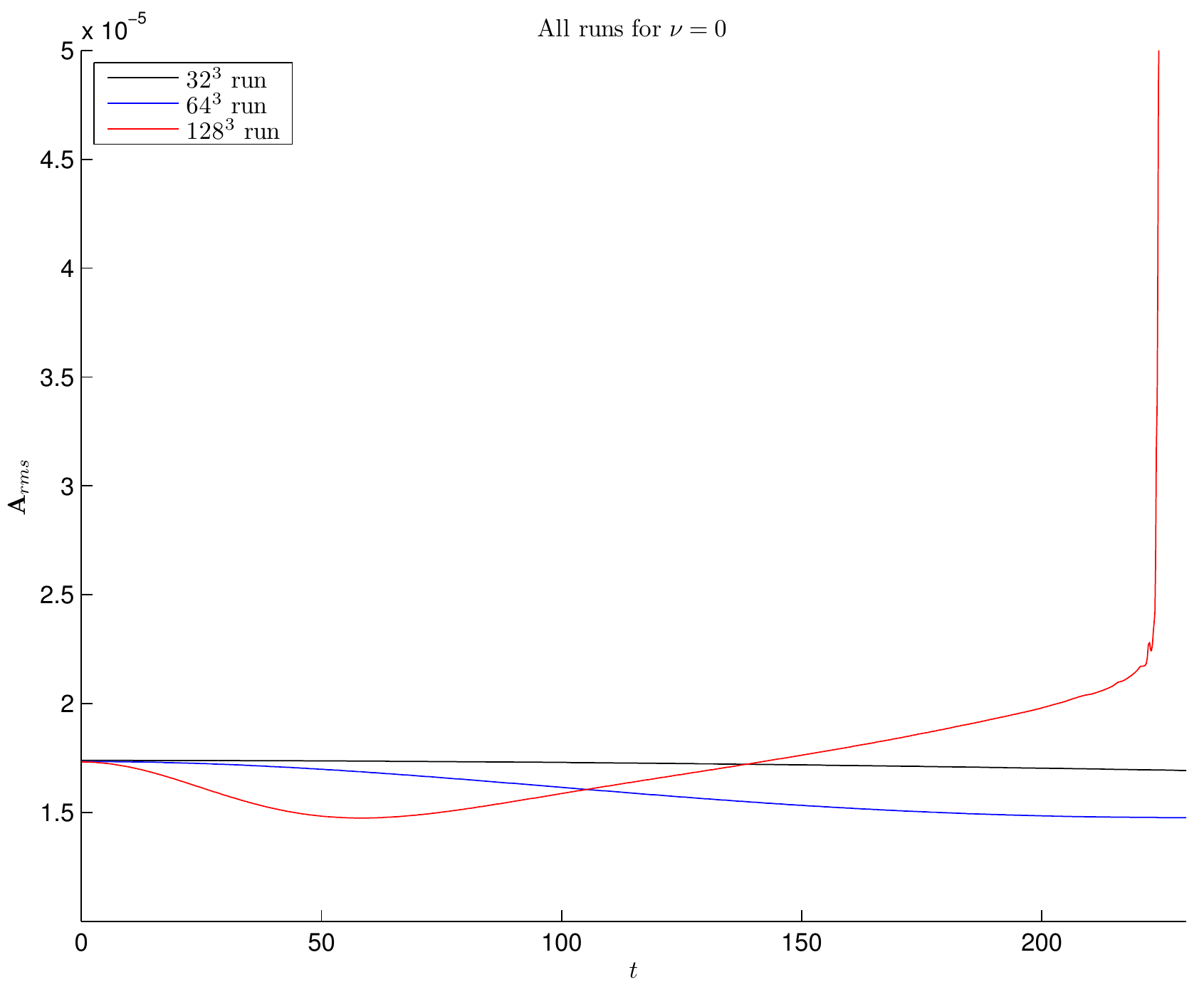}
 \caption{$\eta$=0}
\label{fig:sub1}
\end{subfigure}%
\centering
\begin{subfigure}{.5\textwidth}
  \centering
  \includegraphics[scale=0.4]{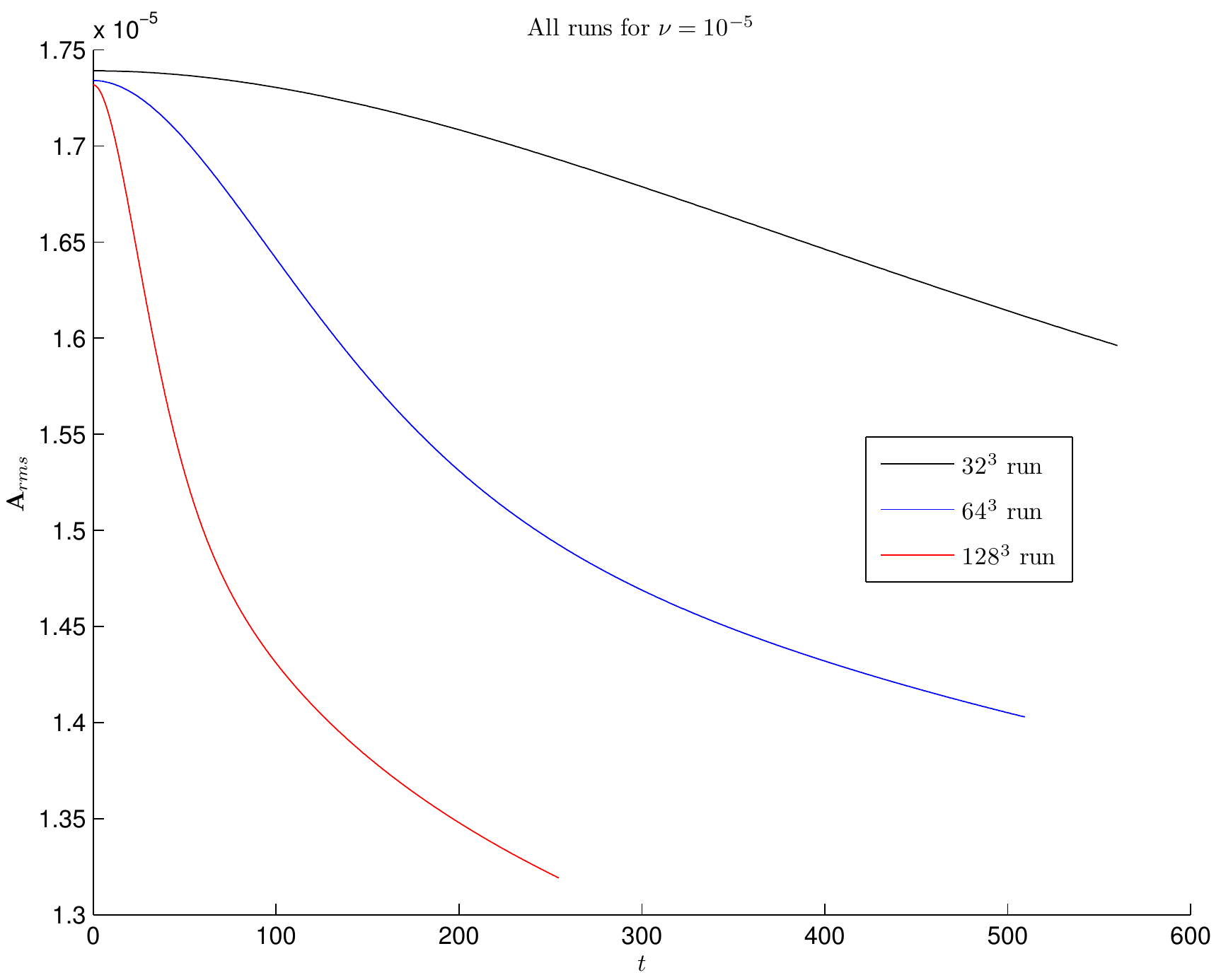}
  
 \caption{$\eta$=0.0001}
\label{fig:sub2}
\end{subfigure}
\\
\centering
\begin{subfigure}{.5\textwidth}
 \centering
 \includegraphics[scale=0.4]{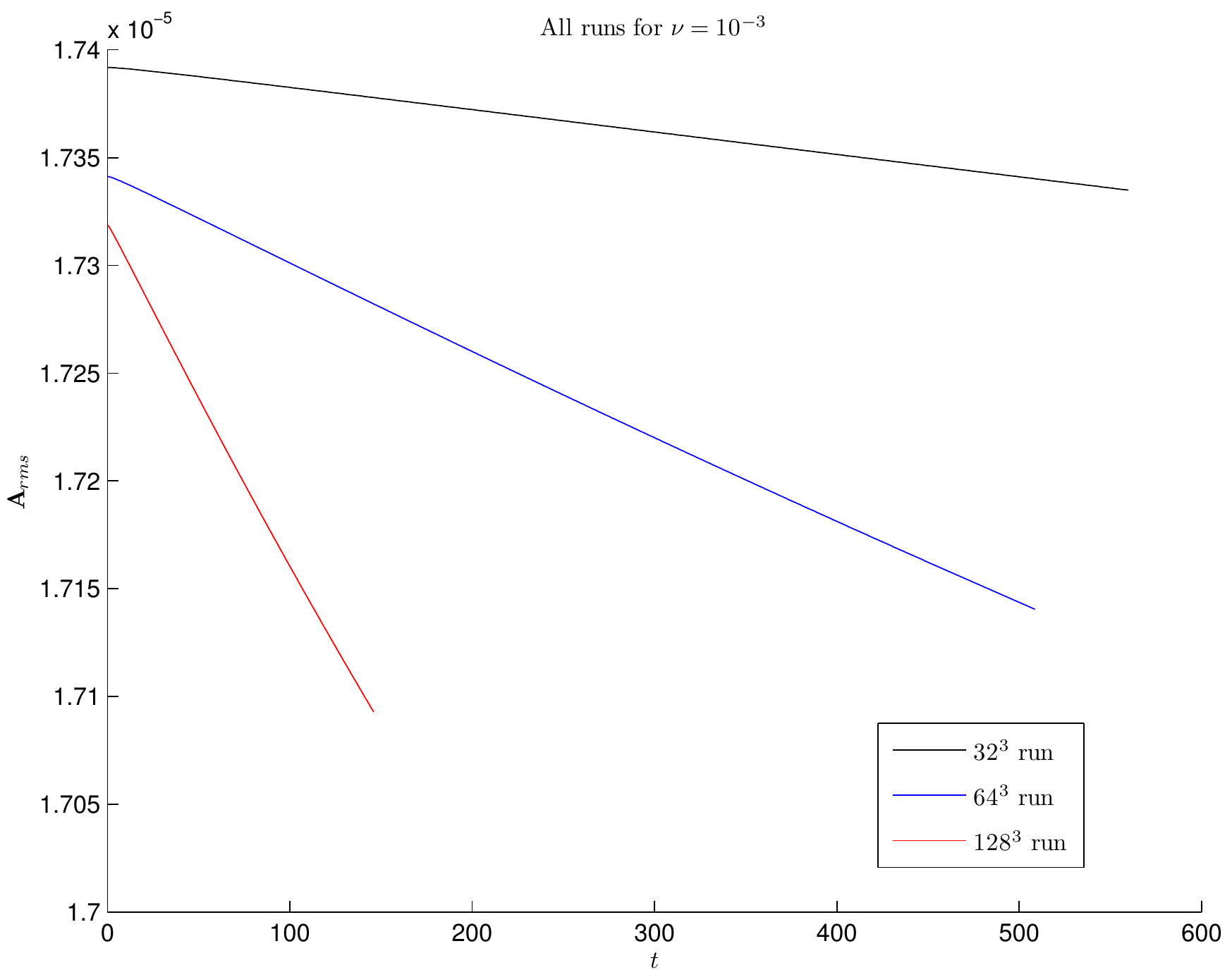}
 \caption{$\eta$=0.0001}
 \label{fig:sub3}
\end{subfigure}%
\centering
\begin{subfigure}{.5\textwidth}
\centering
	\includegraphics[scale=0.4]{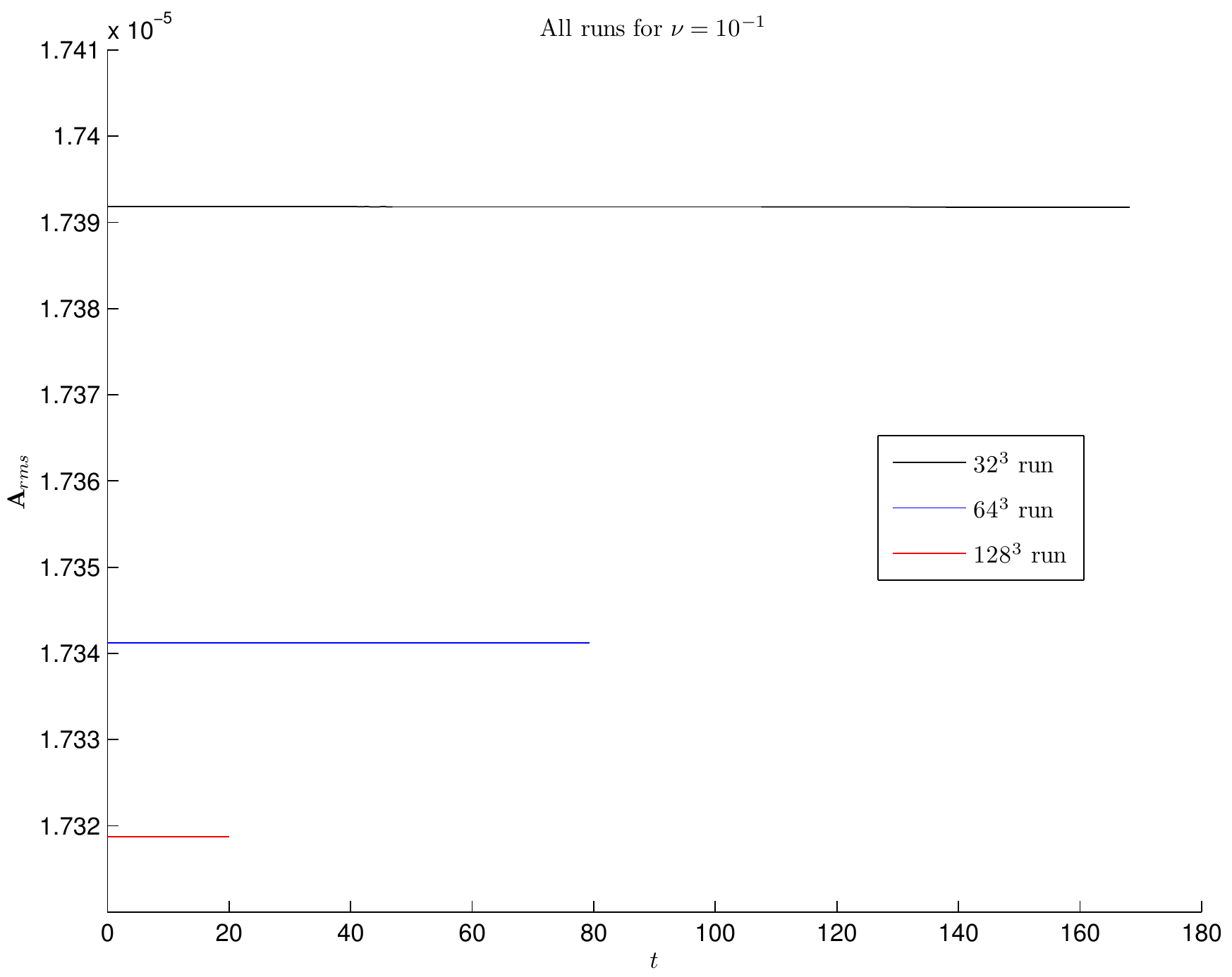}
  \caption{$\eta$=0.1}
\label{fig:sub4}
\end{subfigure}%
\end{figure}

\end{widetext}


\end{document}